\documentclass[11pt]{article}
\usepackage{amssymb,overcite}
\usepackage{multicol,epsfig}
\begin{document}

\title{Phase behaviour of a simple model of globular proteins}

\author{{\bf Richard P. Sear}\\
~\\
Department of Physics, University of Surrey\\
Guildford, Surrey GU2 5XH, United Kingdom\\
email: r.sear@surrey.ac.uk}

\date{}

\maketitle

\begin{abstract}
A simple model of globular proteins which incorporates
anisotropic attractions is proposed. It is closely related to
models used to model simple hydrogen-bonding molecules such as
water. Theories for both the fluid and solid phases are presented,
and phase diagrams calculated. The model protein exhibits
a fluid-fluid transition which is metastable
with respect to the fluid-solid transition for most values of
the model parameters.
This is behaviour often observed for globular proteins.
The model offers an explanation of the difficulty observed
in crystallising some globular proteins and suggests that
some proteins may not have a solid phase at all under all
but extreme conditions.
\end{abstract}

\section{Introduction}

The interactions between globular
proteins are rather poorly understood but it seems clear
that many of the attractive interactions are highly
directional \cite{durbin96,visser92,neal99};
two protein molecules must not only be close
to each other to attract each other but they must also be correctly oriented.
An example is the attraction between hydrophobic patches on
the surfaces of globular proteins; only
if the proteins are oriented so that these parts of their
surfaces face each other is there an attraction.
This attraction is both
highly directional and short ranged \cite{visser92}.
Hydrogen bonds, which are highly directional, are known to have a dramatic
effect on the behaviour of simple molecules such as water.
For example, ice's diamond-like lattice and very low density are imposed by
the tetrahedral arrangement of the hydrogen bonds formed by water molecules.
It is striking that much of the language
used to describe the interactions between globular
proteins \cite{durbin96,visser92,neal99}
is the same as that used in describing hydrogen-bonding
fluids \cite{blanca,nezbeda90,sj}.
Therefore, we believe it is very worthwhile to explore 
the phase behaviour of a simple model of globular proteins which includes
directional attractions. We find phase diagrams
of qualitatively the same form as observed in experiment.
In particular, we find metastable transitions
between dilute and dense fluid phases.

Solutions of globular proteins, like simple molecules, exhibit
two types of phase transition: fluid-to-fluid, analogous
to a vapour-liquid transition, and fluid-to-solid.
The behaviour of globular proteins differs from that
of simple molecules in one important respect: the transition
between a dilute and a dense fluid phase is metastable with respect
to solidification, \cite{broide91,muschol97} whereas,
simple molecules such as argon and water have
equilibrium liquid phases. The proteins crystallise
in a variety of different lattice types. This is suggestive
of directional attractions as at low temperature
these impose the lattice. For example,
a water molecule bonds to four molecules arranged
tetrahedrally around it. In
a diamond lattice but not a face-centred-cubic lattice each water molecule is
surrounded by four others arranged
tetrahedrally, and so water freezes into a solid with a diamond
not a face-centred-cubic lattice.
Isotropic attractions of any range have a relatively weak
effect on the solid phase; the solid is
face-centred cubic as it is for hard spheres.
Globular proteins are not perfectly spherical, of course, and
this may also have an effect on the lattice type.

Previous work \cite{rosenbaum96,tenwolde97,poon97,sear99b} has ignored
any directionality in the attraction and modeled the protein
molecules with a potential consisting of a hard core and a short-range
isotropic attraction. The phase behaviour of such a potential is
qualitatively similar to experiment and to
the behaviour we find for our model,
with the exception of the solid phase.
We will return to this point when we discuss our results in section
5 but the interaction between proteins almost certainly
includes both isotropic and directional attractions \cite{neal99}.

The work is presented as follows. In the next section we specify
the simplest model of a protein we can devise and 
which includes directional attractions.
It does not include isotropic attractions and all the directional
attractions are of the same strength.
This model is very similar
but not identical to very simple models of water
\cite{nezbeda90,ghonasgi93,duda98,vega98}.
Given its simple nature and our lack of knowledge of the
interaction between globular proteins our analysis will be
purely qualitative. Nevertheless, we can describe the experimental
phase behaviour and the phase diagrams we calculate have intrinsic interest
as some of the first calculations which include the fluid and
solid phases when the attractions are highly directional.
For related work see Ref. \citen{vega98}.
The theories for the fluid and solid phases are derived in
sections 3 and 4, respectively.
We restrict ourselves to a brief summary of the theory for
the fluid phase as it is well established,
it is based on the formalism derived by Wertheim 
\cite{wertheim84,wertheim86b,wertheim86,wertheim87,jackson88,chapman88}.
The theory for the solid phase is new, it is a cell theory, like
that of Vega and Monson \cite{vega98}.
Example phase diagrams are shown and discussed in section 5.
We finish with a discussion of the connection between the interactions
and both the fluid-fluid transition and the stability or
otherwise of the solid phase.

\section{Model}

Our model is not new, it is of a type which has been widely employed to model
hydrogen-bonding molecules such as water \cite{nezbeda90} and alcohols
\cite{blanca}. Indeed, the idea of a potential consisting of a
repulsion plus a sticky patch goes back to Boltzmann \cite{boltzmann}.
Specifically, we will consider the conical-site model of
Chapman, Jackson and Gubbins \cite{jackson88,chapman88}.
A schematic of the model is shown in Fig. 1.
The potential is a pair potential $\phi$ which is a sum of two
parts: a hard-sphere repulsion, $\phi_{hs}$, and a set of sites
which mediate short-range, directional attractions. The sites
come in pairs: a site on one particle binds only to its partner
on another particle. So, the number of sites, $n_s$, is constrained
to be an even number. The two sites of a pair are numbered consecutively
so that an odd-numbered site, $i$, binds only to the even-numbered site,
$i+1$. This is the only interaction between the sites, an odd-numbered
site, $i$, does not interact at all with sites other than the $(i+1)$th
site. Each site has a specific orientation, see Fig. 1.
The orientation of site number $i$
is specified by means of a unit vector ${\bf u}_i$.

We can write the interaction potential between a pair of particles as
\begin{equation}
\phi(r_{12},\Omega_1,\Omega_2)=\phi_{hs}(r_{12})
+\sum_i^{'}
\left[  \phi_{ii+1}(r_{12},\Omega_1,\Omega_2),
+  \phi_{ii+1}(r_{12},\Omega_2,\Omega_1)\right],
\label{pot}
\end{equation}
where the dash on the first sum denotes that it is restricted to
odd values of $i$. The interactions between the sites
on the two particles are $\phi_{ii+1}(r_{12},\Omega_1,\Omega_2)$, which
is the interaction between site $i$ on particle 1 and site $i+1$ on
particle 2, and $\phi_{ii+1}(r_{12},\Omega_2,\Omega_1)$, which
is the interaction between site $i$ on particle 2 and site $i+1$ on
particle 1. These are functions of $r_{12}$, $\Omega_1$ and $\Omega_2$,
which are the scalar
distance between the centres of particles 1 and 2, the orientation
of particle 1 and the orientation of particle 2, respectively.
The particle is rigid, but not axially symmetric, so
its position is completely specified by the position of its centre and
its orientation $\Omega$, which may be expressed in terms of the
three Euler angles \cite{allen87}.

The hard-sphere potential, $\phi_{hs}$, is given by
\begin{equation}
\phi_{hs}(r)=
\left\{
\begin{array}{ll}
\infty & ~~~~~~ r \le \sigma\\
0 & ~~~~~~ r > \sigma\\
\end{array}\right. ,
\label{hspot}
\end{equation}
where $\sigma$ is the hard-sphere diameter. The conical-site
interaction potential $\phi_{ii+1}$ is given by \cite{jackson88}
\begin{equation}
\phi_{ii+1}(r_{12},\Omega_1,\Omega_2)=
\left\{
\begin{array}{ll}
-\epsilon & ~~~~~~
r_{12}\le r_c ~~~~~ \mbox{and} ~~~~~ \theta_{1i}\le\theta_c ~~~~~ \mbox{and}
 ~~~~~ \theta_{2i+1}\le\theta_c\\
0 & ~~~~~~ \mbox{otherwise}\\
\end{array}\right. ,
\label{sitepot}
\end{equation}
where $\theta_{1i}$ is the angle
between a line joining the centres of the two particles and the
unit vector ${\bf u}_i$ of particle 1, and
$\theta_{2i+1}$ is the angle
between a line joining the centres of the two particles and the
unit vector ${\bf u}_{i+1}$ of particle 2.
The conical-site potential depends on two parameters:
the range, $r_c$, and the maximum angle at which a bond is
formed, $\theta_c$. Of course, as the attractions are directional,
$\theta_c$ will be small, no more than about $30^{\circ}$.
The attractions are also short ranged, $r_c$ no more than 10\%
larger than $\sigma$.

The angles between the site orientations, the vectors ${\bf u}_i$
will determine which solid lattice is formed. This is like water,
where the diamond-like lattice is enforced by the tetrahedral
arrangements of the hydrogen bonds. For simplicity, we will take the
sites to be arranged such that they are compatible with a simple cubic
lattice. Then if we express the unit vectors ${\bf u}_i$ in
Cartesian coordinates, $(x,y,z)$, then when we have four sites, $n_s=4$,
the set of vectors
${\bf u}_1=(1,0,0)$, ${\bf u}_2=(-1,0,0)$, ${\bf u}_3=(0,1,0)$ and
${\bf u}_4=(0,-1,0)$ would describe our model.
For six sites then we add two additional sites at orientations
${\bf u}_5=(0,0,1)$ and ${\bf u}_6=(0,0,-1)$.
Note that this arrangement of sites is perfect for a cubic lattice,
in general we would expect the angles between the attractions to
be not quite perfectly aligned with the angles in the solid phase,
These mismatches will tend to destabilise the solid phase so
our model is an extreme case; a general arrangement of sites will
tend to have a less stable solid phase.

\section{Theory for the fluid phase}

The theory for the fluid phase of particles interacting via a hard-core
and directional attractions mediated by sites is well established.
Wertheim developed highly accurate perturbation theories for
models with one \cite{wertheim84,wertheim86b}
and two sites \cite{wertheim86,wertheim87}.
By perturbation theory we mean a theory which calculates the
difference in free energy between a fluid of particles
interacting via the full potential and that of a fluid
of particles interacting just via the hard-sphere part of the
potential.
Chapman, Jackson and Gubbins, by assuming additivity
of the effect of the sites on the free energy, generalised the
perturbation theory to potentials with more than two sites
\cite{jackson88,chapman88}.
The theory has been extensively tested against the results
of computer simulation \cite{jackson88,ghonasgi93,duda98,vega98}.

Thus, the perturbation theory gives for the Helmholtz free energy per
particle of the fluid phase, $a_f$, \cite{jackson88}
\begin{equation}
\beta a_f (\eta,T)=\beta a_{hs}(\eta)
+ n_s\left[\ln X+\frac{1}{2}(1-X)\right],
\label{af}
\end{equation}
where $a_{hs}$ is the Helmholtz free energy per particle of a fluid of
hard spheres. We use an accurate expression
derived from the equation for the pressure of Carnahan and
Starling \cite{carnahan69,hansen86} for $a_{hs}$. The volume
fraction $\eta=\rho(\pi/6)\sigma^3$ is a reduced density;
$\rho$ is the number density of particles.
$\beta=1/kT$, where $k$ is Boltzmann's constant and $T$ is the
temperature. $X$ is the fraction of sites which are {\em not}
bonded to another site. As all site-site interactions are equivalent
the fraction of each type of site which is not bonded is the same.
The fraction of sites which are bonded and the fraction
which are not bonded must, of course, add up to one. Thus we can
simply write down a mass-action equation for $X$, \cite{jackson88}
\begin{equation}
1=X+\rho X^2 Kg_{hs}^c(\eta)\exp(\beta\epsilon),
\label{ma}
\end{equation}
where 
$g_{hs}^c$ is the contact value of pair distribution function
of a fluid of hard spheres. It can be obtained from
Carnahan and Starling's expression for the pressure of
hard spheres using the virial equation \cite{hansen86}.
The volume of phase space (both translational and orientational
coordinates) over which a bond exists is $K$ \cite{jackson88},
\begin{equation}
K=\pi\sigma^2(r_c-\sigma)(1-\cos\theta_c)^2.
\label{kdef}
\end{equation}

See Refs.
\citen{wertheim84,wertheim86b,wertheim86,wertheim87,jackson88,chapman88,sear96,sj}
for a detailed discussion and derivation of Eqs. (\ref{af}) and (\ref{ma}).
The rightmost term in Eq. (\ref{ma}) is the fraction of sites which
are bonded. This is equal to the fraction of sites which are unbonded
and so are available to form a bond, $X$, times the
probability of such a site forming a bond. This probability is
the density of other sites
available for bonding, $\rho X$, times a Boltzmann factor
$\exp(\beta\epsilon)$ and integrated over the phase volume for
which a bond exists. The factor of $g_{hs}^c$ is required at high
density to take account of the effect of surrounding particles
pushing particles together.
As for the free energy, Eq. (\ref{af}), it is  easy to check that, for
$n_s=2$, it gives the correct low density limit \cite{sear96}.

The mass-action equation, Eq. (\ref{ma}), is a quadratic equation
for $X$ and the physical solution is
\begin{equation}
X=\frac{2}{1+[1+4\rho Kg_{hs}^c\exp(\beta\epsilon)]^{1/2}},
\label{x}
\end{equation}
Note that Eqs. (\ref{ma}) and
(\ref{x}) are not quite the same as the equivalent
equations in Refs.
\citen{wertheim84,wertheim86b,wertheim86,wertheim87,jackson88,chapman88,sear96,sj}.
In those references $\exp(\beta\epsilon)$ is replaced by
$\exp(\beta\epsilon)-1$. As $\beta\epsilon$ is quite large,
five or more, the difference between the two is very small.
Also, our $K$ is $4\pi$ times the $K_{AB}$ of Ref. \citen{jackson88}.

The pressure of the fluid phase, $p_f$, can be found by differentiating the
free energy, $a_f$:
\begin{eqnarray}
\beta p_f&=&\rho^2\left(\frac{\partial\beta a_f}{\partial\rho}\right)\\
&=&\beta p_{hs}+n_s\rho^2\left(\frac{\partial X}{\partial \rho}\right)
\left(\frac{1}{X}-\frac{1}{2}\right)
\end{eqnarray}
where $p_{hs}$ is the pressure of a fluid of hard spheres,
and the derivative of $X$ may be obtained by taking the
derivative of Eq. (\ref{x}).
The chemical potential of the fluid phase $\mu_f=a_f+p_f/\rho$.
The state of our single component fluid is specified by the ratio of the
site energy to the thermal energy, $\beta\epsilon$,
and the density, $\rho$, or volume fraction, $\eta$.
If there is phase coexistence at any temperature then the two
coexisting densities can be obtained from the
requirement that the chemical potential and pressure be
equal in the two phases. This requirement yields a pair of
coupled nonlinear equations which may be solved to give the two
coexisting densities.

\subsection{The second virial coefficient}

In experiments on proteins, the strength of the attractions
is often assessed by measuring the second virial coefficient, $B_2$,
via a scattering experiment. The second virial coefficient
is equal to the coefficient of the term linear in density
in the Helmholtz free energy per particle \cite{hansen86}.
From Eq. (\ref{af}) we see that it will be the sum of
the second virial coefficient of hard spheres,
$B_2^{hs}=(2\pi/3)\sigma^3$, and a contribution
from the sites. This contribution of the sites
may be obtained by first setting $g_{hs}^c$ to one in Eq. (\ref{x});
one is its value to lowest order in density. Then we
expand out Eq. (\ref{x}) for $X$
as a density expansion and truncate after the linear term.
This truncated density expansion for $X$ is then
substituted in Eq. (\ref{af}), and the result
expanded out in density. The linear term is the second virial
coefficient. The result is
\begin{equation}
B_2= B_2^{hs} - \frac{n_s}{2}K\exp(\beta\epsilon).
\label{b2}
\end{equation}

\section{Theory for the solid phase}

At low temperature, solidification is driven by the attractive interactions,
not packing effects as it is with hard spheres \cite{hoover68} or
with particles with only one or two sites \cite{sear95}.
In the solid phase the position of a particle's centre of mass and its
orientation are constrained by the need for it to always be in such a position
and orientation that none of the bonds it makes with the
neighbouring particles are broken. Like hard spheres the particle
is constrained by its neighbours
but unlike hard spheres the constraint is not only due
to the requirement that the particle not overlap with one of its
neighbours but also
due to the requirement that bonds not be broken,

We will use a cell theory to describe the
free energy of the solid phase of our model \cite{buehler51,sear98}. 
This means we will approximate the energy by the energy when all bonds
are formed, the ground-state energy, and the entropy by the logarithm
of the phase volume available to a particle when it is constrained
to lie in the solid, forming these $n_s$ bonds with its neighbours.
Vega and Monson \cite{vega98} have used a cell theory to describe
the solid phase of a very similar model, a simple model of
water. They avoid a couple of the approximations used here at the
cost of not having an analytical free energy. Comparison with
the results of computer simulation show that their cell theory is
quite accurate.
Within a cell theory for a solid phase,
the Helmholtz free energy per particle, $a_s$, is given by
\begin{equation}
\beta a_s=-\ln q_P
\label{adef}
\end{equation}
where $q_P$ is the partition function of a single particle
trapped in a cage formed by the requirements that all its $n_s$
sites bond to neighbouring particles, and that its hard core
not overlap with any of these neighbours.
In order for these bonds to not be broken the particle
must always be within $r_c$ of the surrounding particles.
This fixes the lattice constant, $a$, at a little less than $r_c$.
It is a little less as when the particle moves off the lattice
site it will be moving towards some of its neighbours and
away from others. Thus it can explore regions where it is further
than $a$ from some of its neighbours.
Of course, a particle cannot move within $\sigma$ of any of its
neighbours due to the hard-sphere interaction. Thus the
particle can move a distance $a-\sigma$
in the direction of any of its neighbours.
The exact value of the maximum lattice constant for which
the particle can move about, constrained by the hard-sphere
interactions, without breaking any bond, is difficult to
estimate; as is the volume available to the centre of mass of
the particle \cite{buehler51}.
Therefore, we approximate the lattice constant $a$ by
$r_c$ and the volume to which the particle is restricted by
$(r_c-\sigma)^3$.
The requirement that no bonds be broken also severely restricts
the orientations of the particle. When a non-axially symmetric
particle is free to rotate it explores an angular phase space of
$8\pi^2$. However, in the solid its rotations will be restricted
to those which are small enough not to violate the requirement
that the orientations of its site vectors are within $\theta_c$
of the lines joining the centre of the particle with the those
of the neighbouring particles. Again the exact value
of angular space available to the particle is complex, and
it also depends on the position of the particle. We
approximate this angular space by assuming that each of the
three angular degrees of freedom can vary independently
over a range of $2\theta_c$. The normalised angular space
available to a particle in the solid phase is then
$(2\theta_c)^3/8\pi^2=\theta_c^3/\pi^2$.
The energy per particle is, of course, $-(n_s/2)\epsilon$, and so
the partition function, $q_P$, is then just the volume available to the
centre of mass of the particle times
the angular space available times $\Lambda^{-1}\exp[(n_s/2)\beta\epsilon]$,
where $\Lambda^{-1}$ is the integral over the momentum degrees of freedom.
Thus, we have for $q_P$,
\begin{equation}
q_P=(r_c-\sigma)^3\left(\frac{\theta_c^3}{\pi^2}\right)
\Lambda^{-1}\exp\left(\frac{n_s}{2}\beta\epsilon\right)
\end{equation}
Inserting this expression for $q_P$ into Eq. (\ref{adef}),
\begin{equation}
\beta a_s=-3\ln\left(\frac{r_c}{\sigma}-1\right)
-\ln\left(\frac{\theta_c^3}{\pi^2}\right)
-\frac{n_s}{2}\beta\epsilon
=\beta \mu_s,
\label{mus}
\end{equation}
where we have neglected a term $\ln(\Lambda/\sigma^3)$.
This is the free energy at a lattice constant of $r_c$.
The maximum possible density of a simple-cubic lattice is
when the lattice constant $a=\sigma$, then the density is
$\sigma^{-3}$.
This density corresponds to a volume fraction $\eta=\pi/6$.
When the lattice constant is $r_c$, the density is  
$r_c^{-3}$ and the volume fraction is
$(\pi/6)(\sigma/r_c)^3$.
On compression, the free energy of solid phase increases very
rapidly and diverges when $a=\sigma$.

We are interested in finding coexistence between the solid phase and
the fluid phase at low temperature, when our assumption that
no bonds are broken in the solid phase will be accurate. Then the
pressure at coexistence will be low and the solid will be near the
its minimum possible density, $r_c^{-3}$.
The chemical potential $\mu_s=a_s+p_s/\rho$ where $p_s$ is the
pressure and $\rho$ is the density. At low pressure $p_s/\rho$
contributes a negligible amount to the chemical potential, which
enables us to equate $a_s$ and $\mu_s$ as we have done in Eq. (\ref{mus}).
The coexisting fluid density at the fluid-solid transition
is then found by equating the chemical potentials in the two phases.
The density of the coexisting solid phase, when the temperature is low enough
that solidification is driven by the attractive interactions
not packing effects, is assumed constant at $r_c^{-3}$.
Essentially, the requirement that the pressure be equal in
the two phases disappears due to the fact that in the solid
phase the pressure is very sensitive to density.
This is best seen in the schematic
diagram of the free energy shown in Fig. 2.
The free energy has very narrow minimum at a density
of $r_c^{-3}$; at higher densities
it rapidly increases as the maximum density of the simple-cubic
lattice is approached and at lower densities it is much higher
as some of the bonds are broken. The pressure (= minus
the slope of the free energy with respect to the volume)
then varies very rapidly. See Fig. 11 of Ref. \citen{vega98} which
shows the results of computer simulations for a similar solid phase;
the pressure varies from almost zero to $20kT/\sigma^{-3}$
over a range in volume fraction of about 0.01.
Now, the more rapid the
variation the smaller the displacement in density that is required
to vary the pressure of the solid phase to equal that in the fluid
phase. By fixing the density of the solid phase to be $r_c^{-3}$
and then determining the density of coexisting fluid phase by
just equating the chemical potentials in the two phases, we
are neglecting this displacement in density.

\section{Phase behaviour}

The only phase transition of hard spheres is a fluid-solid transition
\cite{hoover68}. Spheres which have only one site and so form only pairs
of spheres, also only have a fluid-solid transition \cite{sear95}.
When there are two sites, linear chains of particles form,
again there is only a fluid-solid transition \cite{malanoski97}.
All these transitions are driven by packing effects, hard spheres
pack more efficiently in the solid than in the fluid phase.
When there are more than two sites,
there is a fluid--fluid transition, according to the free energy
of the fluid phase, \cite{notecp} Eq. (\ref{af}).
Also, when there are more than two sites a fluid-solid transition
driven by the interactions between sites not the packing of the hard
cores, is possible. Indeed at low temperatures the sites will
drive the solidification, they will impose the lattice type. It 
will be the one which maximises the number of sites which can interact.
This low temperature solid is well described by the theory of section 4.
We will only plot results at low temperature when this theory
is accurate.
Nezbeda and Iglesias-Silva \cite{nezbeda90} were the first
to observe a fluid--fluid transition in a model of the type
we are considering here.
Their model has four sites and was developed as a simple model of water.

Figures 3 to 5 are example
phase diagrams of four, Figs. 3 and 4, and
six site, Fig. 5, model proteins.
In each case, as
$n_s>2$, the fluid phase free energy predicts a transition
between a dilute and a dense fluid phase.
The values of $\beta\epsilon$, $\eta$ and $B_2$ at
the three critical points,
$(\beta\epsilon)^{cp}$, $\eta^{cp}$ and $B_2^{cp}$, respectively,
are tabulated in Table 1.
However, only
in Fig. 4 does this transition appear in the equilibrium
phase diagram.
In Figs. 3 and 5 the
fluid--fluid transition is metastable. This means that
at equilibrium the transition does not occur, it is
preempted by the fluid-solid transition, but if the
fluid can be cooled sufficiently far into the two-phase
region that it crosses the metastable fluid--fluid
coexistence curve then a fluid--fluid transition might
be observed. Whether this can be done or not depends on the
dynamics of crystallisation \cite{debenedetti,tenwolde97,sear99b}.
A metastable fluid-fluid transition was also predicted
by Vega and Monson \cite{vega98}.
Experiments on globular proteins have found such metastable
fluid--fluid transitions \cite{broide91,muschol97}; the crystallisation
of proteins is often slow, taking several days, which allows
the protein solution to be cooled into a region of the phase
diagram where the fluid phase separates into two fluid phases
of differing densities.

Our toy model of a protein has three parameters:
the number of sites, $n_s$, the range of the attraction
between sites, $r_c$, and the angle over which two sites
interact, $2\theta_c$. Comparing Figs. 3 and 4
we see that decreasing $\theta_c$
destabilises the dense fluid phase (the `liquid')
with respect to the solid phase leaving only a fluid-solid
transition in the phase diagram. Comparing Figs. 4
and 5 we see that increasing the number of sites has the
same effect. 
Increasing the range of attraction has a weaker effect,
although our theory predicts that increasing the range
actually tends to drive a fluid-fluid transition metastable.
This is opposite to what is found for
isotropic attractions where increasing the range
stabilises the dense fluid phase, resulting in the appearance
of an equilibrium fluid-fluid transition,

The particles have six degrees of freedom,
three translational and three rotational.
In order for a particle to form a bond with another
particle both its position and orientation must be within
small bounds. The distance between the centre of mass of the
particle and that of the particle to which it is bonding must
lie within the range $\sigma$ to $r_c$. The orientation
of the site vector ${\bf u}_i$ must also lie within $\theta_c$
of the line joining the two centres of mass.
From Eq. (\ref{kdef}) we see that in order for a bond to
be formed between a pair of particles one of the
translational degrees of freedom must be restricted
to the small range $r_c-\sigma$ (the other two correspond to
motion over a sphere of radius $\simeq\sigma$ and so are much
less restricted), and two of the orientational degrees
of freedom must be restricted (the other one corresponds to
rotation about the axis of the bond and is unrestricted).
In the solid phase all six degrees of freedom are restricted
to small ranges, $\simeq r_c-\sigma$ for the translational
degrees of freedom and $\simeq \theta_c$ for the orientational
degrees of freedom. However, in the solid phase the particle
forms $n_s$ bonds. In the fluid phase each bond that forms
requires the restriction of one translational and two
orientational degrees of freedom.
So, in the solid phase when we go from four to six sites
then the entropy does not change by much as in both
cases all six degrees of freedom are restricted. Our
theory, Eq. (\ref{mus}), actually predicts that the entropy
is identical for four and for six sites but this is a
consequence of the approximations we have made.
But of course on adding two extra sites the energy per particle
decreases by $\epsilon$. Thus the extra sites reduce the chemical
potential of the solid phase, this then coexists with a lower density
fluid phase: a fluid phase at a density below the lower
of the two fluid-phase densities of the fluid-fluid transition.
The fluid-fluid transition is then metastable.

On examining Table 1 we see that neither the critical volume fraction,
$\eta^{cp}$, nor the value of the second virial coefficient at the critical
temperature, $B_2^{cp}$, are sensitive to the value of $\theta_c$.
However, $\eta^{cp}$
is sensitive and $B_2^{cp}$ is very sensitive to the number of sites. 
Broide {\it et al.} \cite{broide91} estimate that the volume fraction
of the globular protein $\gamma$-crystallin
in the solution at the critical point is
approximately 20\%, whereas that of lyzozyme is approximately 15\%.
To give some idea of what we might expect $\eta^{cp}$
to be, its value in the van-der-Waals fluid is 0.13.
The van-der-Waals fluid is a fluid
of hard spheres with a very long-range attraction \cite{hansen86,sear95}.
Both our six-site model and short-range isotropic attractions
yield critical volume fractions above 0.13 but our four-site model
predicts a lower volume fraction.
Thus, the experiments rule out modeling the attractions as
just four sites, at least for $\gamma$-crystallin.
This is not a surprise, we expect there to be
an isotropic attraction present and are neglecting it here purely
for simplicity.

As for the second virial coefficient at the critical temperature.
It is equal to $-3.5\sigma^3$ for a van-der-Waals fluid.
For short-range isotropic attractions it
is less negative than this \cite{hagen94}. Indeed as the range of
an isotropic attraction tends to zero the fluid-fluid
transition crosses the fluid-solid transition and becomes a
solid-solid transition and the value
of the second virial coefficient at the critical point of
the solid-solid transition tends to towards its value for hard spheres,
$(2/3)\pi\sigma^3$ \cite{bolhuis94}.
Correspondingly, $\eta^{cp}$ tends
towards the close-packing volume fraction for hard spheres, $\simeq0.74$.
The second virial coefficient of solutions of globular
proteins can be measured via a scattering experiment.
Measuring it under conditions near those
for which the metastable fluid-fluid transition has its critical
point would shed some light on the nature of the interactions.
For instance, a large and negative second virial coefficient would
rule out isotropic attractions as the dominant cause of the
fluid-fluid transition. A small and negative
second virial coefficient would however be ambiguous;
both short-range isotropic attractions and our six-site
model would be consistent with this finding.

Our approximation for the free energy of the solid phase, Eq.
(\ref{mus}) is quite crude. We can estimate the accuracy of the
phase diagrams we have calculated by replacing our approximation
of the lattice constant $a$ by $r_c$ which overestimates $a$, by
the more conservative estimate $a=(r_c+\sigma)/2$
\cite{vega98}. With this approximation the trends we observe
do not change however the range of parameter values over which the
fluid-fluid transition is metastable does change. For example,
with the parameter values used in Fig. 3 and approximating
$a$ by $(r_c+\sigma)/2$ the fluid-fluid transition is just stable.
It is stable for values of $\theta_c\ge0.28$ whereas with
the approximation $a=r_c$ the fluid-fluid transition is only
stable for $\theta_c\ge0.43$. By stable we mean that the critical
point of the fluid-fluid transition does not lie within the
fluid-solid coexistence region. The approximation
$a=(r_c+\sigma)/2$ also, of course, implies a denser solid
phase than $a=r_c$.

\section{Conclusion}

A simple model of a globular protein molecule, incorporating
highly directional interactions, has been proposed.
Example phase diagrams, Figs. 3 to 5,
have been calculated. As observed by Vega and Monson \cite{vega98}
highly directional attractions can have phase diagrams which are
similar to those of short-range isotropic attractions; there
is a fluid-fluid transition but it is metastable. We have also been
able to show that if the number of sites is not too large,
then directional attractive interactions
can induce a stable fluid-fluid transition, see Fig. 4.
As far as we are aware this is the first demonstration which shows
a fluid-fluid transition driven by
directional attractions which is definitely stable with
respect to a fluid-solid transition. There have been several
calculations of fluid-fluid transitions driven entirely by directional
attractions with no isotropic attraction present
\cite{nezbeda90,ghonasgi93,duda98} but only Vega and Monson \cite{vega98}
considered the solid phase.

Clearly, our potential with only directional attractions is as much
a caricature of the interaction between globular proteins as is
a potential with only a short-range isotropic attraction.
It does have one advantage, however, in that, unlike isotropic
attractions, the solid phase is not necessarily either
face-centred-cubic or hexagonal-close-packed.
In our model the lattice is determined by the
arrangement of the sites. We chose an arrangement which
allowed all four or six sites to form bonds in a simple-cubic
lattice but we could equally well choose arrangements of sites
which can bond in, say, a body-centred-cubic lattice.
Globular proteins form solids with a variety of different
solid lattices.
If the interaction between globular proteins can be represented
without too much error by a pairwise additive potential then
the limited amount that we know of interactions between globular
proteins suggests that
a potential with a hard-core, and both isotropic and directional
attractions is the most appropriate one to choose.
Such potentials have been used with considerable success to predict
the phase behaviour of hydrogen-bonding liquids \cite{blanca,gilvillegas97}.

Our model does raise one issue: that of differences between the many
globular proteins. Previous work
\cite{rosenbaum96,tenwolde97,poon97,sear99b}
on models with isotropic attractions
inherently assumes that all globular proteins are alike. This
supposition is supported by the presence of metastable fluid-fluid
phase transitions for several different globular proteins. However,
the ease with which a globular protein crystallises differs
enormously from protein to protein. Indeed the fact that some
have never been crystallised provides much of the motivation behind
the work. If we vary the orientations of the site vectors, ${\bf u}_i$,
then we dramatically change the stability of the solid phase. If the
vectors are arranged in way which is not compatible with any solid
lattice then at low temperatures there will be no solid phase (except
at extremely high osmotic pressures). If in the solid phase not
all the bonds can be made but they can in the fluid phase then the
fluid will not solidify as this will increase the energy, and at
low temperature the energetics dominates.
We speculate that the
proteins which have resisted efforts to crystallise them may have
directional attractions which are not compatible with any solid lattice.
If this is so then crystallising them will be all but impossible
without minimising the strength of these directional attractions
which will act to stabilise the fluid phase.

\newpage

\begin{table}
\begin{tabular}{|c|c|c|c|}
\hline
Figure & $(\beta\epsilon)^{cp}$ & $\eta^{cp}$ & $B_2^{cp}$\\
\hline
3 & 10.24 & 0.090 & $-15.5\sigma^3$ \\
4 & 8.64 & 0.090 & $-15.5\sigma^3$\\
5 & 7.18 & 0.154 & $-4.0\sigma^3$\\
\hline
\end{tabular}
\end{table}
\vspace*{0.1in}

\noindent
Table 1. The ratio of the interaction site energy to the thermal energy
at the critical temperature, $(\beta\epsilon)^{cp}$,
the second virial coefficient
at the critical temperature, $B_2^{cp}$, and the volume fraction at the
critical point, $\eta^{cp}$, are tabulated for the parameters
used in Figs. 3 to 5. Both $\eta^{cp}$ and $B_2^{cp}$
are identical for the parameters of Figs. 3 and 4
as within our theory for the fluid phase changing either $\theta_c$
or $r_c$ simply changes the temperature scale of the phase behaviour.

\vspace*{0.3in}
\noindent
Fig. 1. A schematic diagram of our model of a globular protein.
The circle represents the hard core of the particle and
the shaded circles represent three of the four conical
attraction sites (the fourth is on the other side of the sphere).

\vspace*{0.3in}
\noindent
Fig. 2 A schematic diagram of the free energy of coexisting fluid and
solid phases. The solid curves are the free energy per particle
(over $kT$). The dashed line joins the coexisting phases.

\vspace*{0.3in}
\noindent
Fig. 3. The phase diagram of our model of a globular protein. The number
of sites $n_s=4$, $r_c=1.05\sigma$ and $\theta_c=0.3$ radians
or about 17$^\circ$.
The solid curves separate the one and two-phase regions.
The letters F, S and 2 denote the regions of the phase space
occupied by the fluid phase, the solid phase and coexistence
between the fluid and solid phases. The dashed
curve is the coexistence curve for a metastable fluid--fluid
transition.

\vspace*{0.3in}
\noindent
Fig. 4. The phase diagram of our model of a globular protein. The number
of sites $n_s=4$, $r_c=1.05\sigma$ and $\theta_c=0.45$ radians
or about 26$^\circ$.
The solid curves separate the one and two-phase regions.
The letters F, S and 2 denote the regions of the phase space
occupied by the fluid phase, the solid phase and coexistence
between two phases, either two fluid phases or
the fluid and solid phases. The dashed
curves are the metastable continuation of the fluid--fluid
coexistence curve, and the dotted curve is the
metastable continuation of the high-density-fluid--solid
coexistence curve. The latter terminates when it touches
the spinodal of the fluid-fluid transition.

\vspace*{0.3in}
\noindent
Fig. 5. The phase diagram of our model of a globular protein. The number
of sites $n_s=6$, $r_c=1.05\sigma$ and $\theta_c=0.45$ radians
or about 26$^\circ$.
The solid curves separate the one and two-phase regions.
The letters F, S and 2 denote the regions of the phase space
occupied by the fluid phase, the solid phase and coexistence
between the fluid and solid phases. The dashed
curve is the coexistence curve for a metastable fluid--fluid
transition.

\newpage
\begin{figure}
\begin{center}
\epsfig{file=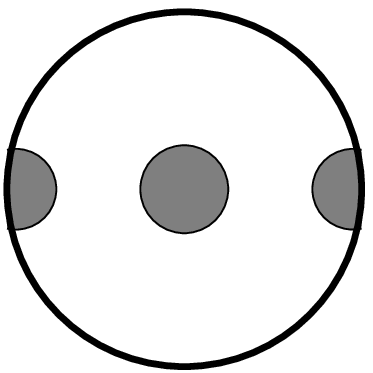,width=2.5in}
\end{center}
\end{figure}
~~\\

\newpage
\begin{figure}
\begin{center}
\epsfig{file=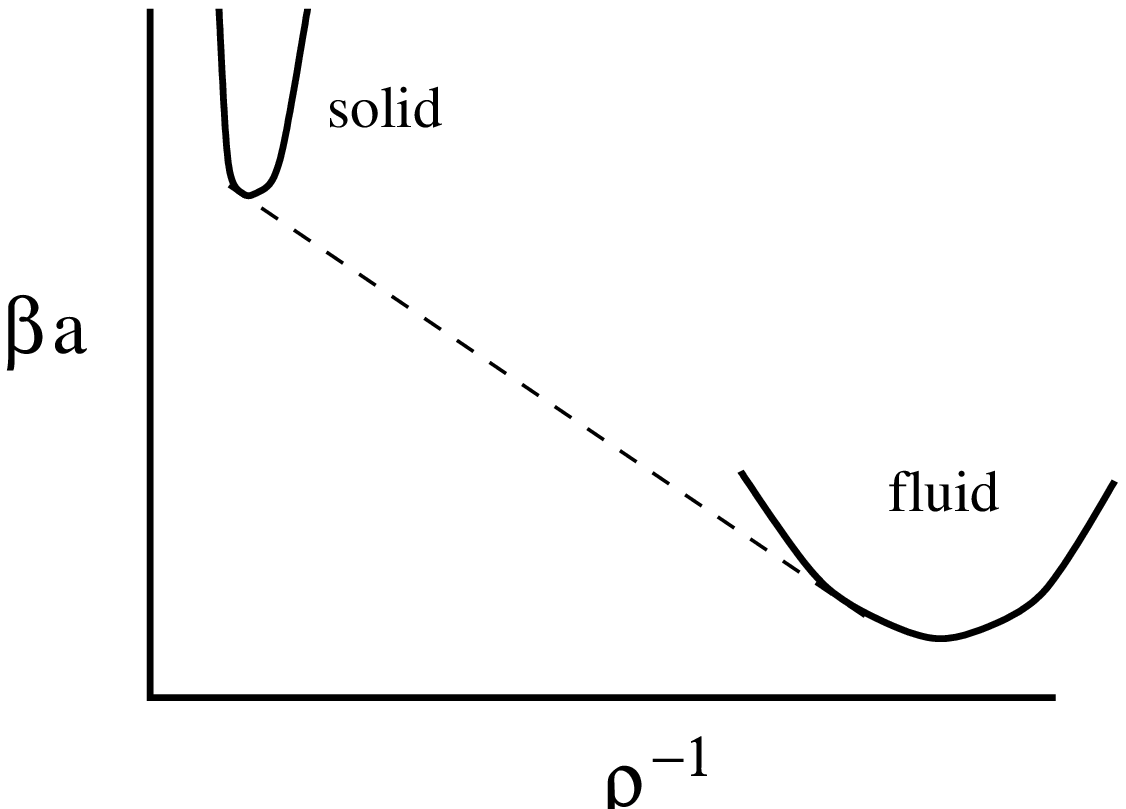,width=3.5in}
\end{center}
\end{figure}
~~\\

\newpage
\begin{figure}
\begin{center}
\epsfig{file=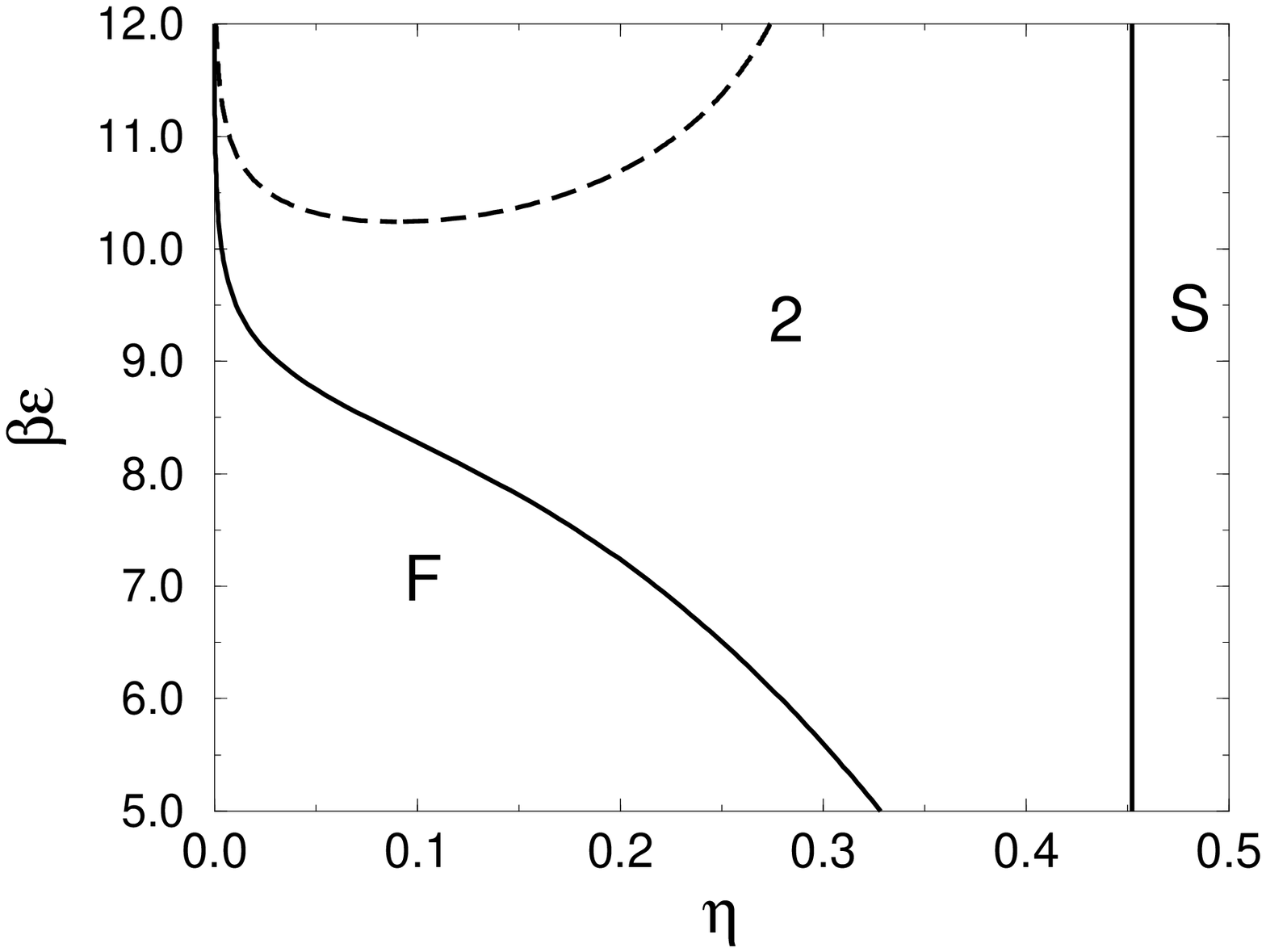,width=3.5in}
\end{center}
\end{figure}
~~\\

\newpage
\begin{figure}
\begin{center}
\epsfig{file=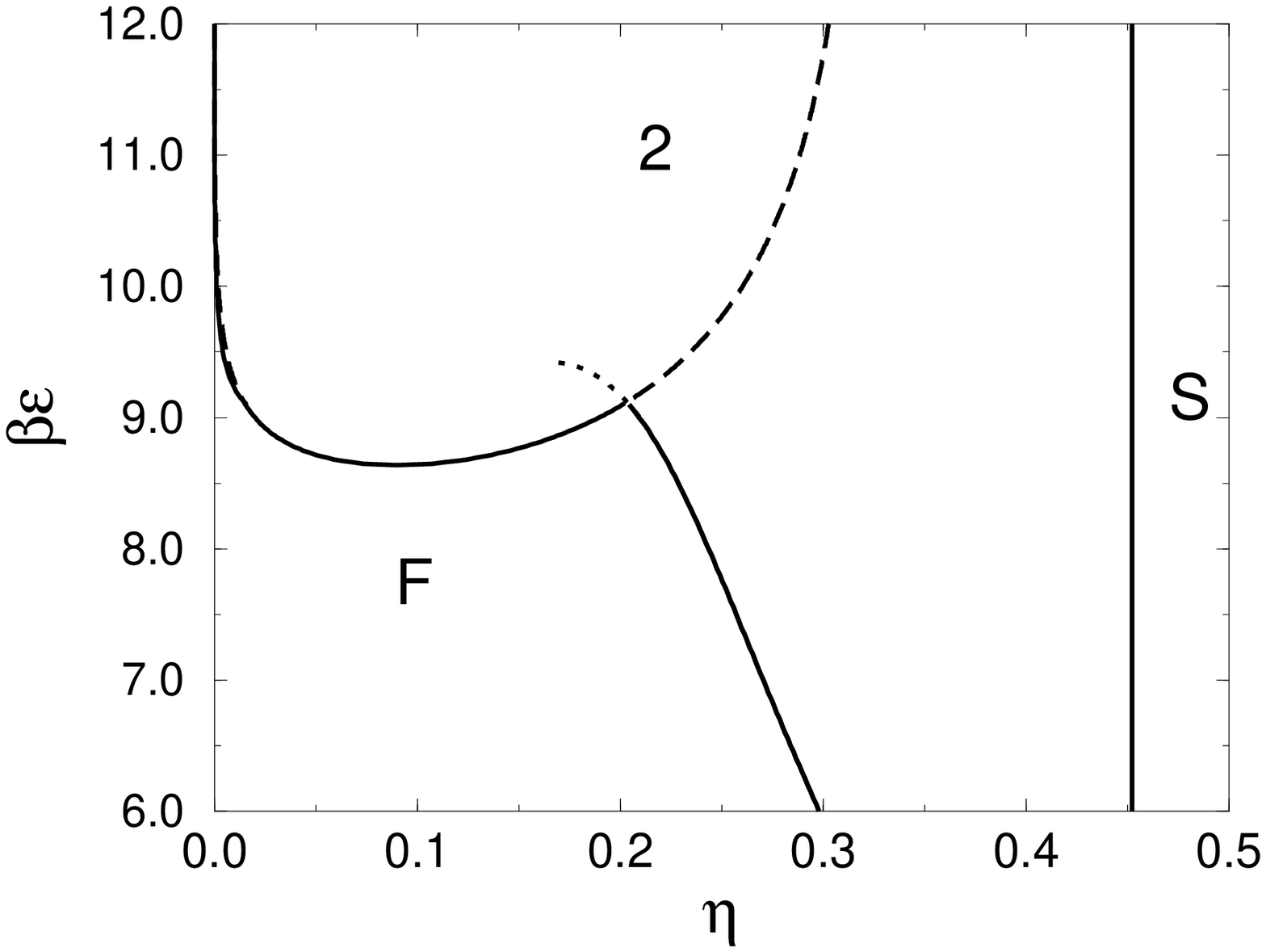,width=3.5in}
\end{center}
\end{figure}
~~\\

\newpage
\begin{figure}
\begin{center}
\epsfig{file=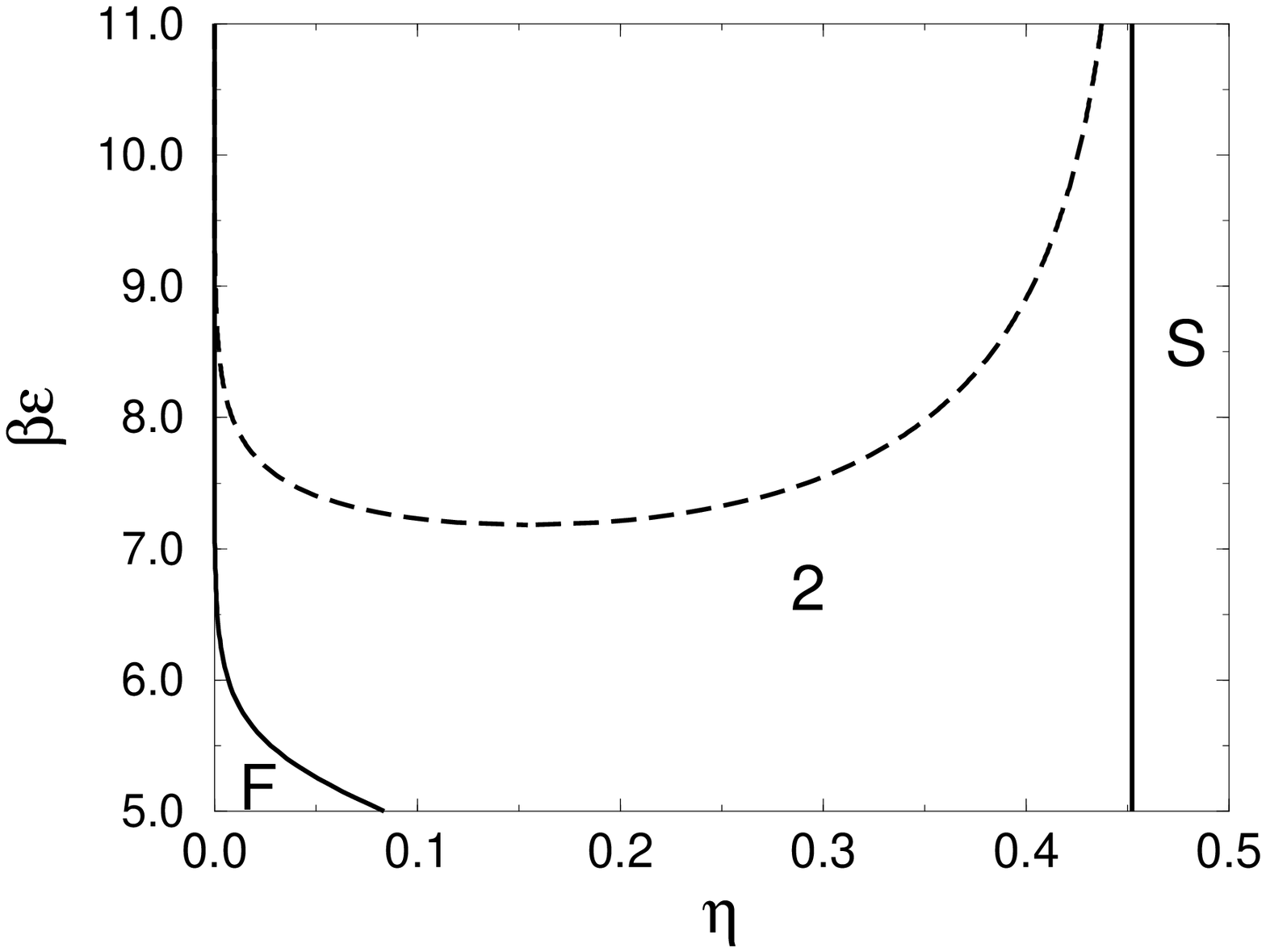,width=3.5in}
\end{center}
\end{figure}
~~\\


\begin{thebibliography}{99}
\bibitem{notecp} Although our model is only defined for $n_s$
an even integer, in the free energy of Eq. (\ref{af}),
we can treat $n_s$ as a continuous variable. If we do so we find
a fluid-fluid transition for $n_s\ge2$. As $n_s\rightarrow2$
the temperature and density at the critical point of the
fluid-fluid transition both tend towards zero \cite{duda98}.
\end{thebibliography}
\end{document}